\documentclass{elsart}
\usepackage{epsfig} \usepackage{changebar} \usepackage{amsmath}

\newcommand{\be}{\begin{equation}}
\newcommand{\ee}{\end{equation}}
\newcommand{\bea}{\begin{eqnarray}}
\newcommand{\eea}{\end{eqnarray}}

\usepackage{amssymb}

\begin{document}

\begin{frontmatter}



\title{Simulations of Large-scale WiFi-based Wireless Networks:
 Interdisciplinary Challenges and Applications}

 \author[label1,label2]{Maziar Nekovee and}
  \author[label2]{Radhika S. Saksena}
 \address[label1]{Mobility Research Centre, British Telecom, Polaris 134, 
Adastral Park, Martlesham, Suffolk IP5 3RE, UK }
\address[label2]{Centre for Computational Science, University College London, 20Gordon Street, London WC1H 0AJ, UK}



\begin{abstract}
Wireless Fidelity (WiFi) is the fastest growing wireless technology to date.
In addition to providing wire-free connectivity to the Internet 
WiFi technology also enables mobile devices to connect 
directly  to each other and form highly dynamic wireless adhoc networks.
Such distributed networks can be used to perform cooperative  
communication tasks such ad data routing and information dissemination in 
the absence of a fixed infrastructure. Furthermore, adhoc grids 
composed of wirelessly networked 
portable devices are emerging as a new paradigm in 
grid computing.
In this paper we review computational and algorithmic challenges of 
high-fidelity simulations of such WiFi-based wireless communication and
computing networks,
including scalable topology maintenance, mobility modelling,
parallelisation and synchronisation. We explore 
similarities and differences between  the simulations of these 
networks and simulations of interacting many-particle systems, such as 
molecular dynamics (MD) simulations.
We show how the cell linked-list algorithm which we have 
adapted  from our MD simulations can be used 
to greatly improve the computational performance of wireless network 
simulators 
in the presence of mobility, and illustrate with an example from our 
simulation studies of worm attacks on mobile wireless adhoc networks. 

\end{abstract}
\begin{keyword}
Wireless Computer Networks \sep Large-scale Simulations 
\sep Interacting Many-Particle Systems \sep Wireless Grids
\PACS 
\end{keyword}
\end{frontmatter}

\section{Introduction}
Modern world has become increasingly mobile. As a result, traditional 
ways of connecting users to the Internet (and to each other) 
via physical cables have proved inadequate. Wireless communications
\cite{wicom}, on the other hand,  poses no restrictions on the user's mobility and allows a great deal 
of flexibility, both  on the part of users and service providers. 
Wireless connectivity for 
voice via mobile telephony made it possible for people to connect to each 
other regardless of location. This  has had a profound influence on
the business of telecommunications, as well as the society as a whole
\cite{economist1}. New  wireless technologies targeted at computer networks promise to do the same for Internet access, connecting wirelessly not only laptops and portable devices but also  millions of cars, sensors, consumer devices, etc to each other and 
to the global Internet.

The most successful, and fastest growing, example of such wireless 
technologies is WiFi (Wireless Fidelity)\cite{wifi}.
Like cellular technology, WiFi uses a number of base stations to 
connect user devices to an existing fixed network (these base stations
are called access points). However, unlike cellular
systems which are centralised, WiFi systems operate in a highly
distributed fashion. Each WiFi device is responsible for 
managing its own connectivity, mobility and access to the radio spectrum.
Furthermore, unlike cellular systems, nearby WiFi devices can directly connect
to each other and form  self-organising wireless adhoc
networks \cite{adhoc1, adhoc2}. Such networks are highly dynamic and flexible. 
They can be created (and torn down) 
on the fly in order to route data packets between participating devices, 
or to the closest Internet gateway. Adhoc technology can also be used to 
connect together a collection of WiFi accesspoints which then form a so-called 
mesh network \cite{mesh}.
This can help to greatly extend the range of WiFi coverage without 
the need for connecting every single accesspoint to the fixed network.

Initially WiFi technology was used 
to provide connectivity to ``nomadic'' users in coffeshops, 
airports etc, and for wire-free Internet access in homes and offices.
The last few years, however, have seen the emergence of much more ambitious 
applications of this technology.
For example, it is expected that WiFi-based 
wireless access will enable the coverage of entire cities, thus providing 
citizens on the move with high-speed (11-54 megabits per second) 
connectivity. Other frontiers in WiFi technology include 
high-speed Internet access to automobile users, 
and WiFi-based vehicular adhoc networks and vehicular grids 
\cite{maziar-vgrid}. 

In addition to the above applications in telecommunications, wireless 
adhoc grids based on WiFi and related technologies are emerging as a new 
paradigm in grid computing \cite{wgrid1,wgrid2}. Adhoc grid environments 
enable mobile users to join together wirelessly and share computing resources, 
services and information \cite{wgrid2}. One example  of such  adhoc grids 
are wireless sensor grids for medical, industrial and environmental 
monitoring. Another one are wireless computational 
grids where WiFi-enabled devices are networked 
together in order to perform complex computing and data aggregation tasks, 
which are beyond the capabilities of a single device
\cite{maziar-vgrid}.
 
The increasing complexity and the very large scale of such emerging WiFi
systems has created a need for scalable high-fidelity simulation  platforms that can help scientists, engineers and network planners accurately predict and 
optimise their performance prior to large-scale deployment.
The aim of this  paper is to review computational challenges that are 
involved in creating  such simulation platforms, including 
scalable topology construction and mobility, parallel and distributed 
simulations on grid platforms, 
and  synchronisation. We also show that there are 
interesting similarities between the simulations of 
WiFi-based wireless networks and molecular dynamics (MD) simulations of interacting many-particle systems, and illustrate 
how these could be exploited in practice.

The rest of this paper is organised as follows.  
In section 2 we give a brief description of the main 
ingredients for simulations of WiFi-based
networks. This is followed by an examination of the computational and 
algorithmic challenges  of such simulations, and how to 
address these. Section 4 describes, as an example,
aspects of our simulation studies  of computer worm attacks on 
mobile wireless adhoc networks. We close the paper in section 5 with 
conclusions.

\label{intro}

\section{Modelling Ingredients}
\label{models}
There has been significant previous research in modelling \cite{internet} and 
simulations of wired communication networks \cite{ns2,riley}. 
However, modelling of wireless networks is very distinct from modelling of wired networks in that the physical medium 
properties, i.e. radio propagation and interference, cannot be separated from the higher layer 
network  protocols, because strong interactions impact performance and drive 
engineering design decisions. Furthermore the ability 
of users to (rapidly) change their physical location while maintaining 
connectivity greatly increases the dynamism of these networks, in comparison to fixed networks. In this section we shall focus on describing 
these distinctive ingredients for  the modelling of WiFi-based wireless networks. Other components in high-fidelity simulations of these networks, which are 
outside the scope of the current paper include modelling of 
various packet routing mechanisms \cite{wicom,routing} and  transport protocols
\cite{tcp} in WiFi environments.

Building on 
the idea of creating parallels between these networks and interacting
many-particle systems, we first consider 
the simplest building block of such networks, namely a single pair of
communicating  WiFi devices. We then consider many 
such pairs operating in the vicinity of each other, and  discuss the 
interaction topologies  and operational rules of the 
resulting systems.

\subsection{A single pair of WiFi devices}
At present WiFi  devices are constrained by regulators to operate in pre-defined
frequency bands. Depending on the IEEE standard used in the device these 
frequency bands are either in the 2.4-2.5 GHz or in the
5.2-5.8 GHz range. Each of these bands is divided into a 
number of frequency channels.  

Consider two WiFi devices $i$ and $j$ which communicate with each
other  using a common frequency channel, $f_i$. The received radio signal 
strength at device $j$ resulting from  a transmission by device $i$
depends on  a variety of effects. These include free space attenuation of 
radio waves, the response of the environment and mobility; 
the latter will be neglected for clarity of presentation but will 
be picked up in section 2.5. Effects due to the environment include
reflection at surfaces, diffraction due to obstacles, and transmissions 
through walls. Phenomenologically, in the absence of detailed information on 
the environment, these effects can be  described using the so-called pathloss 
model \cite{pathloss},  which states that signal power at a receiving device $j$ is related 
to the signal power of the transmitting node $i$  via the following 
equation:
\be
P^{ij}=\frac{P^i}{c_{f_i}r_{ij}^\alpha}.
\ee
In the above equation 
$r_{ij}$ is the distance between node $i$ and node $j$, 
$P^{i}$ and $P^{ij}$ are the transmit power and the received power, 
respectively, and $c_{f_i}$ is a frequency-dependent 
constant. 
For free space propagation $\alpha=2$, but depending on the specific 
indoor/outdoor propagation scenario it is found empirically 
that this exponent can vary
typically between $2$ and $5$. A data transmission by node $i$ is correctly 
received at node $j$, i.e. $i$ can establish a communication link with 
$j$, provided that: 
\be
\frac{P^{ij}}{\nu}= \frac{P^i/c_{f_i}r_{ij}^\alpha}{\nu} \ge \beta_{th}.  
\ee
In the above equation 
$\beta_{th}$ is a sensitivity threshold and $\nu$ is the noise level at 
node $j$.

Condition (2) translates into a maximum transmission range for node $i$:
\be
r_t^i= \left( \frac{P^i}{c_{f_i} \beta_{th} \nu} \right)^{1/\alpha},
\ee
such that device $i$  can establish a wireless link with device $j$
only if $j$ is within a circle of radius $r_t^i$. 

\subsection{A collection of WiFi devices: Interference effects}
In the above we considered a stand-alone sender-receiver 
pair of WiFi devices. In reality, however, many pairs of nearby 
wireless nodes may simultaneously attempt to establish links, 
either between themselves or to 
nearby access points.  Due to the broadcast nature of radio transmissions, 
a radio signal transmitted towards a specific node can interfere with communications of many nearby network nodes and contribute to their noise 
level. Consequently, a successful transmission from 
node $i$ to node $j$ depends not only on the transmit power of node $i$ and 
its distance to node $j$ but also on the activity of all other 
nearby nodes. 
In particular, {\it aggregate} transmission
of nearby devices may result in a situation where a transmitter-receiver pair 
cannot establish a link despite the fact that they are within the  range of each other.
Such interference effects need to be accurately taken into account in 
modelling data communications in wireless networks in order 
to obtain realistic results \cite{bogodia1}.

To model interference, for each signal transmitted 
from a sender $i$ to a receiver $j$  
the aggregate received power resulting from all other  nearby sender 
devices needs to be computed. Signal arrival at node $j$ is then 
considered successful only if  the ratio of the 
received power from $i$  to aggregate noise is above the sensitivity 
threshold, $\beta_{th}$. Computing the impact of interference on the transmissions of devices 
is one of the most computationally expensive components in the  simulations
of wireless networks as it requires the computation of $ O(N^2)$ pairwise 
interactions. However, taking advantage of the fact that the 
interference effect decays as $1/r^\alpha$ with the distance between two 
devices, one usually limits the computation of the interference terms 
to devices which are within a so-called interference range of  a given 
device. Conventionally the interference range is chosen to be 
$r_i^j=2r_t^j$\footnote{The precise choice of the interference range 
depends on both the decay exponent $\alpha$ and the density of 
nodes in the system and may have to be increased significantly 
in order to obtain accurate results \cite{bogodia2}}.

\subsection{Medium Access Control (MAC)}
The interference problem is perhaps the most important issue 
in deployment of high-density WiFi
networks. WiFi technology attempts to mitigate  this problem 
using  a distributed random access mechanisms called Medium Access Control
(MAC). 
The MAC protocol used by  WiFi-based wireless devices follows the 
IEEE 802.11 standard \cite{wifi}, which specifies a set of
rules that enable nearby devices to coordinate their transmissions 
in a distributed manner,
in such a way that devices whose radio transmissions may interfere with each 
other do not get access to the same frequency channel at the same time. 

The IEEE 802.11 MAC is a   
complex protocol and we do not attempt to describe here 
the full model of this protocol.
Instead we focus on the most relevant aspect of this protocol, the so-called 
listen-before-talk (LBT) rule. This rule  dictates that each device should 
check the occupancy of the wireless medium before starting a data transmission 
and refrain for a {\it random} time from  
transmitting if it senses that the medium is busy.
Roughly speaking, the net effect of this mechanism is 
to create an interference-free ``exclusion zone'' around each transmitting 
device (which is roughly of the size $\sim \pi (2r^i_t)^2$), 
thereby reducing (but not entirely eliminating) the possibility of packet 
collisions. 
The presence of the MAC introduces novel spatio-temporal correlations 
in the dynamics of data communications which need to be taken into account 
in realistic simulations of these networks \cite{maziar-worms}.

\subsection{Graph representation of interactions in WiFi Networks}
From the above models of radio propagation, an abstract 
communication graph  for a collection of WiFi devices can be constructed.
This is achieved by creating an edge between node $i$ and all other 
nodes in the plane that are  within the transmission range of $i$, 
and repeating this procedure for all nodes in the network.
In general wireless devices may use different transmit powers 
such that the existence of a wireless link from $i$ to $j$ does not imply 
that a link from $j$ to $i$ also exists. Consequently the resulting 
communication graph is {\it directed}. 
Assuming, however, that all devices use the same transmit power $P$, 
and a corresponding transmission range $r_t$,
the topology of the resulting network can be described as a 
two dimensional random geometric graph (RGG) \cite{rgg1}.
Similarly, one constructs an interference graph for the network by 
creating an edge between any two nodes which are within a radius $r_i$ of each other. Fig 1. shows, as an examples, the communication and 
interference graphs created by a collection of WiFi devices distributed 
randomly in a $1000\times 1000$ $m^2$ rectangular 
area.

Mathematically, a graph is represented by a corresponding  
{\it adjacency matrix} $A$, where the element $a_{ij}=1$ if two nodes are connected, and zero otherwise. Since the adjacency matrices corresponding to the communication and interference graphs 
of WiFi networks are usually  
sparse, they can be efficiently encoded in the computer memory in 
the form of {\it neighbour lists}.

\begin{figure}[t]
 \epsfig{file=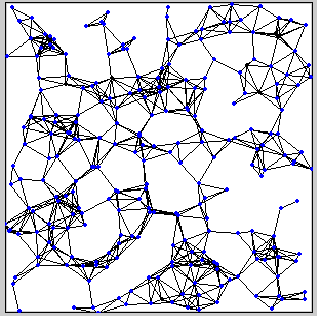, width=2.5in, angle=-90, clip=1} 
 \epsfig{file=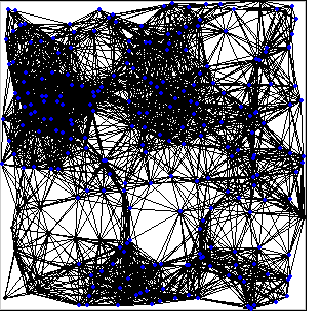, width=2.5in, angle=-90, clip=1} 

 \caption{The communication (left panel) and the corresponding
 interference graph (right panel) of a wireless adhoc network created
 by WiFi devices uniformly distributed in a $1000 \times 1000 $ $m^2$ 
rectangular area.}
\end{figure}

\subsection{ Modelling User Mobility}
The movement of users carrying  WiFi devices 
can greatly influence the performance of WiFi systems.
The impact of user mobility is twofold. First of all, as a result of mobility,
devices  continuously move in and out of each others' transmission and/or
interference ranges and this will  result in {\it time-dependent} network topologies. Secondly, mobility causes spatio-temporal variations in the user density, 
and consequently, the traffic load offered to each access point.

Accurately modelling  user mobility is therefore of great importance in 
high-fidelity simulations of WiFi system. Several 
mobility models have been presented in recent simulation
studies of WiFi and Bluetooth-based mobile wireless networks. The most 
widely used of such models assume random and uncorrelated 
movements of individual devices. These include the random walk model 
and variations thereof, such as the random-waypoint model \cite{camp}.
However, such simple models are unable to reproduce important 
features of user mobility patterns, which result from a 
combination of  correlations \cite{helbig}, environmental constraints
\cite{helbig} and social interactions between users \cite{mascolo}.

Fortunately there has been much 
previous research on agent-based realistic modelling of both human and 
vehicular mobility \cite{helbig,ball}, and such models  
can be coupled to network simulators in order 
to examine in detail the impact of user mobility.
Very recently, for example, we used 
microscopic car-following models to investigate various properties 
of vehicular adhoc networks operating in realistic highway traffic scenarios \cite{maziar-vanet}.

A computational issue in coupling high-fidelity 
mobility simulators to wireless network simulators is the 
large difference in the time scales of the two types of simulations.
For example, car-following models typically 
update the position of vehicles every second. On the other hand, a typical 
timestep of wireless network simulations is $\sim 1$ $\mu s$.
To ensure accurate results the combined simulations of 
mobility and wireless communication should be performed using the smallest
time-step in the problem. In the above example, this means a 
$10^6$ increase in the update frequency of the vehicular traffic simulator
\cite{vanet}.

\section{Computational Challenges}

\subsection{Network topology construction and maintenance}
One of the most computationally intensive portion of 
the simulations of WiFi networks is 
the construction of the neighbour lists which encode the topology 
of the network.
In static networks these neighbour lists can be constructed once and for all 
at the beginning of each simulation. In networks comprising  
mobile nodes, the neighbour lists need to be updated with 
every update of  nodes' positions. Consequently,
optimisation of neighbour list construction algorithm 
becomes critical to the performance of the simulation code.

A brute force implementation of the neighbour list construction involves 
checking the distance of each node from
all the other nodes in the system in order to determine its neighbours.
This approach involves a 
double nested loop iteration over all the nodes in the system and scales 
as O($N^2$), where, $N$ is the number of nodes in the network. The 
computation becomes very expensive as we 
go to larger networks, higher node densities and to networks with 
highly dynamic nodes (such as vehicular networks).

A similar issue is faced in simulations of interacting many-particle 
systems such as molecular dynamics simulations, where, 
for each particle in the system, the interaction forces with the remaining 
particles need  to be calculated to simulate its dynamics. Analogous to 
the transmission/interference range in WiFi networks, in MD simulations
of liquids with short-range interaction potentials, the interaction
force computation is truncated at a cut-off radius, $r_c$. When $r_c$ is 
equal to or smaller than one-third of the linear dimension of the 
simulation box, a cell-linked list method is often adopted
which brings down the force computations from an O($N^2$) calculation 
to O($N$) (in this case N is the number of particles in the many-particle 
simulation system). 

\begin{figure}
\centering
\epsfig{file=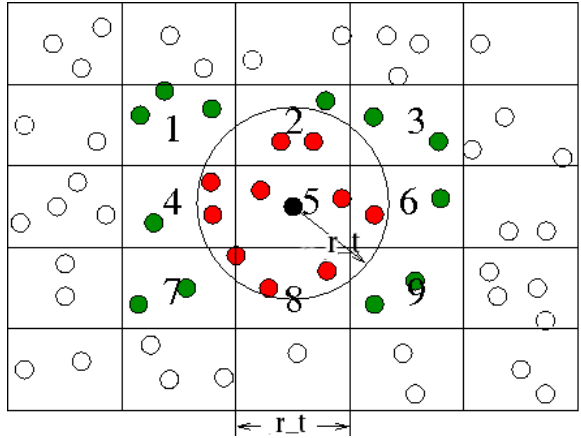, width=3.0in}
\caption{Figure illustrating the cell linked list method 
\cite{allenandtildesley} 
applied to a simulation of WiFi networks. The network
area is sub-divided into sub-cells with a linear dimension equal to the the 
transmission range, $r_t$. For the central node in cell 5, only the other 
occupants in its own sub-cell and those in the immediate neighbouring sub 
cells (1,2,3,4,6,7,8,9) feature in its neighbour list.}
\label{f:sys}
\end{figure}

The cell-linked list approach \cite{allenandtildesley} applied to the 
construction of  network topology in WiFi systems works as follows.
First the two-dimensional network area/simulation 
cell is divided into sub-cells such that the linear dimension of each 
sub-cell is equal to the transmission (interference) range. 
Any node in a sub-cell 
can only interact with nodes in its own sub-cell and in the 
immediate neighbouring sub-cells and is invisible to nodes in all other 
sub-cells in the network. This is illustrated in Figure \ref{f:sys}, 
where, in order to determine the neighbour list for the central node in 
cell 5, one only needs to consider nodes  within cell 5 and in the 
immediate neighbouring cells (1,2,3,4,6,7,8,9). The brute force method of 
constructing neighbour lists with two nested loops over all particles in 
the system is replaced by 1) a loop over all nodes to determine which 
sub-cells they lie in - an $O(N)$ operation and 2) a loop over all nodes to find their neighbour nodes in the immediate neighbouring sub-cells - 
an $O(N\times N_c$) 
algorithm, where $N_c$ is the average number of nodes per sub-cell. 

We note that the above  decomposition of the simulation system into linked cells is naturally suited to domain decomposition parallelisation of the simulations, 
which we shall discuss in the next section. 

\subsection{Parallel and distributed simulations on grid platforms}
Due to the short-range nature of  wireless communications, parallel 
simulations of WiFi-based systems on massively parallel computers or 
tightly coupled grid platforms can be performed most effectively 
using domain decomposition. The area in which WiFi devices operate is 
divided into a number of regular sub-domains with dimensions larger than the 
maximum interference radius of the wireless devices that comprise the 
network. The entire communication stack of all devices within
each subdomain is then allocated to one processor and inter-processor 
communications are only performed when nodes move from one processor to 
another, or there are radio signal propagation across sub-domain 
boundaries. 

Several parallel simulators tools for WiFi-based mobile wireless networks 
which exploit the domain decomposition strategy have been proposed and 
implemented in recent years \cite{glomo,qualnet,vanet,conserve1}. However,
there is very limited published work in examining the 
performance of these simulators in the context of 
large-scale parallel simulations.
Benchmark studies  performed on relatively small  number of processors ($6-12$ 
PEs) however, indicate that only sub-linear parallel speedups can be achieved,
presumably due to a combination of communications and synchronisation overheads.

In addition to domain-decomposition, a task farming strategy can be exploited 
in simulations of large-scale wireless networks in 
order to perform Monte Carlo runs over an ensemble of  network realisations 
and/or to explore the performance for a range of system parameters.
In this case multiple runs of the same code are spawned 
on a set of slave  processors and the results are collected and further processed by the master processor at the end of computation. Unlike domain decomposition, task farming requires no synchronisation and very limited interprocess 
communications. Therefore linear parallel speedups can be achieved 
even for simulations performed on loosely coupled grid platforms.

\subsection{Synchronisation} 
A WiFi network consists of a number of devices each having 
its own internal set of states 
(e.g. the number of data packets in the incoming 
queue of a laptop, or the random backoff time of the MAC protocol). These states 
change {\it stochastically} (and therefore asynchronously)
in response to events which are generated either 
internally or due to interactions with other devices in the system.
For example, the arrival  of a voice call will change the 
state of the outgoing data queue of a WiFi-enabled mobile phone. 

Parallel simulations of interacting systems  with such 
asynchronous dynamics (also known as Parallel Discrete Event
Simulations) requires the use of a synchronisation protocol among 
the processing elements (PES) in order to ensure that 
causality errors are not introduced in the simulation results. 
An  overview of the synchronisation of parallel discrete event 
simulation and a comprehensive discussion of commonly used synchronisation 
schemes can be found in \cite{fujimoto}. 
Conservative synchronisation schemes are conventionally used 
in parallel simulations of  wireless networks \cite{bogodia2,conserve2}.
Each PE defines its own lookahead as the minimum duration 
(measured using the simulation clock) for which it will not send any 
message on its outgoing links. Periodically, a global minimum of all 
PE's simulation time plus their lookahead values is computed.
Each PE can then process all events inside its own domain that take 
place within this time window safely without the need of additional 
synchronisation.

When the number of processing elements 
becomes large, conservative synchronisation schemes may result in  large 
fluctuations in the rates at which different  PES progress,  
hence greatly reducing the computational scalability of the parallel 
discrete event simulation \cite{synch2}.  It has been demonstrated recently  
\cite{synch2} that by changing the  communication topology of PES from 
a regular grid to a {small-world-type} topology, the above problem can be 
eliminated  and high parallel efficiency achieved.
We are currently in the process of implementing such schemes in the 
parallel version of our wireless network simulator and the results 
will be reported elsewhere.

\section{Case Study: Simulations of Worm Attacks on Mobile 
Wireless Adhoc Networks}
Worms are self-replicating computer viruses which can propagate through 
computer networks without any human intervention \cite{book-virus}.
With wireless networks becoming increasingly popular, many 
security experts predict that these networks will soon be a main 
target of attacks by worms and other type of malware \cite{sci-am,nature}.
In addition to individual devices, open resource sites in wireless information 
or computational grids could well be the next wave of targets for such 
wireless worm attacks. A qualitative understanding of such attacks is of great 
importance, both in assessing their risk and for the 
design of effective detection and prevention strategies. 

Worm and virus attacks on the Internet have been the subject of extensive 
empirical, theoretical and simulation studies, and there have been a 
number of studies on securing  conventional wired grids against such  attacks
\cite{wisec}.  Investigation of virus spreading in 
wireless networks and wireless grids in general and worms 
in particular is, however, at its infancy. 
In a recent study \cite{maziar-worms} 
we used Monte Carlo simulations to investigate 
the spreading of worm epidemics in {\it static} wireless adhoc networks.
These studies point out to important differences 
between the propagation patterns of worms in wired and wireless networks
and highligh the importance of incorporating interference effects, 
network topology and medium access control for realistic 
modelling of data communications in these networks.

In this section we briefly describe aspects of these simulations 
and further advance them by incorporating the impact of device 
mobility in the dynamics of worm propagation. 

\subsection{Worm propagation model}
Following \cite{maziar-worms} we assume that wireless 
worms primarily utilise multihop forwarding for their propagation in adhoc
 networks, a mechanism which does not require any Internet connectivity.  
With respect to an attacking  worm we use the so-called 
susceptible-infected-removed (SIR) model from mathematical 
biology \cite{bio}, adapted to the context of wireless communications.
We assume that nodes in the network to 
be in one of the  following three states: vulnerable,
infected, or  immune. Infected nodes try to broadcast the worm to their
neighbours at every possible opportunity.
Vulnerable nodes can become infected with probability  
$\lambda$ when they receive a transmission containing a
copy  of the worm  from an infected neighbour. Finally, infected 
nodes get patched and become immune to the worm with 
probability  $\delta$. We denote by $S(t)$, $I(t)$ and $R(t)$ the 
population of vulnerable, infected and immune nodes, respectively.

\subsection{Simulation details}
We simulated the propagation of worms in mobile 
wireless adhoc 
networks comprising $N$  devices in a $L^2=1000\times 1000$ $m^2$ 
area. At the start of each simulation the devices were spread randomly 
in the simulation area, after which they 
were allowed to move following simple random walks (Periodic 
boundary conditions were used at the edges of the simulation area).
The worm spreading dynamics was 
simulated on top of the resulting time-dependent 
network using  Monte Carlo simulations. Each Monte Carlo run starts
by infecting a single randomly chosen node and proceeds for a certain 
number of simulation timesteps, $i_{update}$, 
after which  the positions of the nodes are updated 
according to the random walk model. We use this form of updating in 
order to mimic the difference in the timescales between the spreading process 
and the user mobility. Each simulation continues  in this fashion 
until the epidemic dies out (i.e. no infected node is left in the network).
We typically average our results over $500$ Monte Carlo runs in order
to obtain statistically significant data. 
Furthermore, the results were also averaged over simulations
starting from different initial infected seeds. 

\subsection{Results}
\begin{figure}
\centering
\epsfig{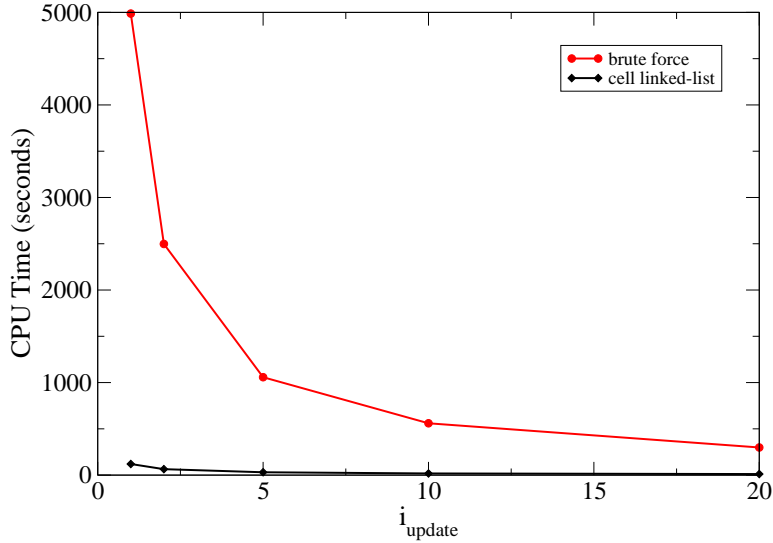}
\caption{Scaling of the computational cost vs. inverse update frequency for 
the simulation of an adhoc network composed of dynamic 
nodes whose positions are updated periodically with frequency $\nu$.
The CPU times are compared for two simulations in which, respectively, 
the brute force method and the cell linked list method were used.}
\label{f:move}
\end{figure}

First we consider the improvement in computational performance 
gained from using the cell linked-list algorithm for updating the 
network topology. 
A comparison of the performance of the cell-linked list and brute force
method for different movement 
update frequencies, $\nu=1/i_{update}$, is shown in Figure \ref{f:move}.
It can be seen that using the cell-linked list algorithm greatly 
reduces the computational cost associated with updating network topology and, 
as expected, this reduction becomes more significant as the update frequency 
increases. 

\begin{figure}
\centering
\epsfig{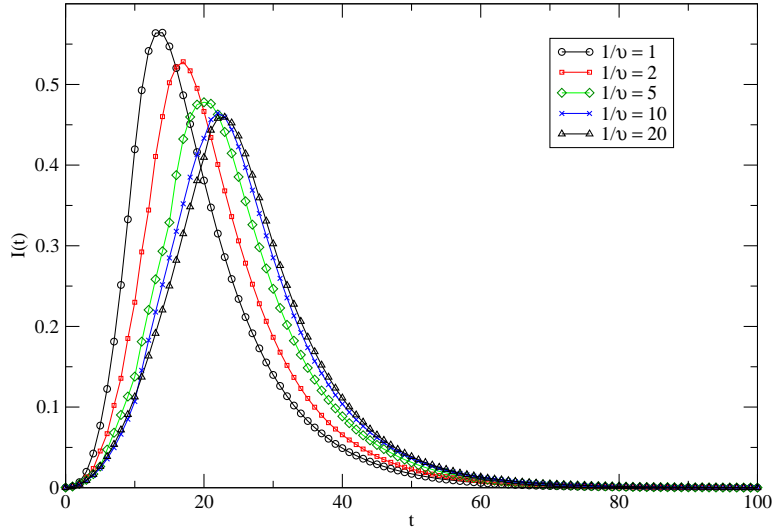}
\caption{Population of infected nodes vs. simulation time in an adhoc network
composed of dynamic nodes whose positions are updated periodically with frequency
$\nu$. The evolution of the infected node population for different update
frequencies is plotted.}
\label{f:it}
\end{figure}

Next we consider the impact of node mobility on the dynamics of worm 
propagation in the network. As an example, 
the time evolution of the population of infected nodes, $I(t)$, is
plotted in Figure \ref{f:it}. The results were obtained for mobile 
wireless adhoc network composed of $4000$ nodes and 
using  $\lambda=0.3$ and $\delta=0.1$. 
The nodes' positions were updated every $1$, $2$, $5$, $10$ and $20$ 
time-steps during 
the worm propagation, i.e $\frac 1 \nu = 1, 2, 5, 10, 20$. It can be seen 
that node mobility has a significant impact on the spreading dynamics.
In particular, as mobility increases (i.e. the network is updated 
more frequently) the epidemic peak (the maximum number of 
infected devices) becomes more pronounced 
and also occurs at earlier times. These results 
indicate that dynamic adhoc networks are  more vulnerable to  
worm attacks than static networks. 

Qualitatively, we can explain the above behaviour in the following way. 
In a fixed wireless adhoc network the  maximum number of nodes to 
which an infected device can spread the worm in the course of its 
infection is limited by the 
total number of devices which are within its transmission range. On average 
this is given by $nA$, where $n$ is device density and $A=\pi r_t^2$ is 
the total area covered by device's transmission. Switching on mobility enables 
infected devices to sweep on a larger area than $A$,
hence increasing the maximum  number of nodes that  each device can infect 
before getting patched. Consequently, both the speed and the magnitude of the 
epidemic are increased with increased device mobility.

\section{Conclusions}
WiFi and related technologies not only allow  users to access the Internet 
on the move, they are also enabling mobile devices to connect directly 
to each other and  form adhoc networks for distributed 
voice, video and data communication. Such adhoc networks also form a flexible 
communication backbone for wireless adhoc grids, an emerging paradigm in 
distributed and grid computing.
With the proliferation of WiFi-enabled mobile devices such grids will 
enable innovative applications based on sharing and federating 
computing and information resources of billions of wireless devices such as
sensors, smartphones, PDAS and laptops.

In this paper we described  aspects of computational modelling 
of such WiFi-based wireless networks, and examined some of the important 
computational challenges which are involved in high-fidelity  
simulations of these networks. 
We argued  that viewing these networks in terms of interacting many-particle 
systems provides a useful framework for understanding and addressing some 
of these challenges. We demonstrated this point by using the cell-linked list 
technique, which is widely used in simulations of such systems, 
for scalable updating of network topology in 
large-scale simulations of mobile wireless adhoc networks. Furthermore, we discussed parallel simulations of WiFi-based wireless networks and 
showed  that conventional parallelisation  strategies  used in parallel
simulations of interacting many-particle systems, such as domain decomposition, are 
readily applicable  to such simulations.  
However, these strategies need to be complimented with 
scalable inter-processor synchronisation schemes in order to deal with 
the asynchronous nature of interactions in wireless networks. 

High-fidelity simulation platforms for  
WiFi-based wireless networks capable of 
effectively  utilising the 
power of computational grid and massively parallel computing are currently at their 
infancy. Such platforms, however, will be  necessary in 
planning  and optimising the highly complex next generation   
WiFi networks. They are also important in realistically analysing the 
potentials and challenges of future adhoc grid platforms
\cite{wgrid1,wgrid2}, such as security, scalability, and  
intermittent network connectivity resulting from mobility.

Our experience 
shows that in designing such platforms one can greatly benefit from 
computational and algorithmic techniques developed in other
branches of computational science. In addition to scalable  
network topology construction and 
parallelisation, which were discussed in the current paper, another 
example that comes to mind is the use of
fast multipole expansion techniques for $O(N)$ interference 
computation in wireless networks with long-range radio signal propagation
\cite{mmp}. At the same time, high-fidelity simulations of WiFi-based 
wireless networks presents an array of new computational challenges 
which are the subject of our ongoing and future research. 

\label{conclusions}
\section*{Acknowledgements}
M.N. acknowledges support from the Royal Society
through an Industry Fellowship. He thanks the Centre for 
Computational Science at UCL for hospitality and his colleagues at BT's
Mobility Research Centre for stimulating discussions.




\begin{thebibliography}{40}
\bibitem{wicom}
A. Goldsmith {\it Wireless Communications}, Cambridge University Press, 2005.

\bibitem{economist1}
{\it Nomads at last: A special report on mobile telecoms}, The Economist,
1-18, April 12th, 2008.

\bibitem{wifi}
M. S. Gast {\it 802.11 Wireless Networks}, Second Edition, O'Reily, 2005;
W. ~Stalling, {\em Wireless Communications Networks}, Prentice Hall, 2005.

\bibitem{adhoc1}
R. Hekmat, {\em Adhoc networks: Fundamental properties and network 
topologies}, Springer, 2006.

\bibitem{adhoc2}
D. Grigoras and M. Riordan, Future Generation Computer Systems {\bf 23},
990-996, 2007.
\bibitem{mesh}
Y. Zhang, J. Luo, H. Hou (eds.), {\it Wireless Mesh Networking: 
Architectures,Protocols and Standards}, Auerbach Publications, NY, 2007.

\bibitem{maziar-vgrid} M. Nekovee, Proc. Workshop on  Ubiquitious Computing and e-Research, 
Edinburgh, UK, May 2005.
\bibitem{wgrid1}
L. W. McKnight, J. Howison, S. Bradner, Wireless grids: Distributed resource 
sharing by mobile, nomadic, and fixed devices, IEEE Internet Computing,
Special Issue on Wireless grids, 24-31, July/August 2004.

\bibitem{wgrid2}
R. Moreno-Vozmediano, A hybrid mechanism for resource/service discovery in 
ad-hoc grids, Future Generation Computer Systems, 2008 (article in press).

\bibitem{ns2}
The Network Simulator -- ns-2, www.isi.edu/nsnam/ns.
\bibitem{riley}
G. Riley, Proc. of the 2003 ACM Winter Simulation Conference, 2003.
\bibitem{internet}
R. Pastor-Satorras and A. Vespignani, {\it Evolution and Structure of the Internet: A Statistical Physics Approach}, Cambridge University Press, 2006.

\bibitem{tcp}
U. O. Ibom, Proc. of the IEEE Information Networking, 1-5, 2008.

\bibitem{routing}
J. J. Garcia, Future Generation Computer Systems, 81-93,1988.

\bibitem{pathloss}
T. Rappaport, {\it Wireless Communications, Principle and Parctice}, 
Prentice-Hall, 2000.

\bibitem{bogodia1}
M. Takai, J. Martin, and R. Bagrodia, Proc. of ACM MOBIHOC'01, 87-94, 2001.

\bibitem{bogodia2}
Z. Ji, J. Zhou, M. Takai, J. Martin, and R. Bagrodia, Proc. of the 18th Workshop on Parallel and Distributed Simulation (PASDs'04). 

\bibitem{rgg1}
M. Penrose, {\it Random Geometric Graphs}, Oxford University Press,
Oxford, 2003.

\bibitem{camp}
See, e.g., T. Camp, J. Boleng, and V. Davis, Wireless Communication and Mobile Computing,
Special Issue on Mobile Adhoc Networking: Research, Trends and Applications
{\bf 2}, 483-502, 2002.


\bibitem{mascolo}
M. Musolesi and C. Mascolo, Mobile Computing and Communications Review {\bf 1}, 
1-12,2006. 

\bibitem {ball} P. Ball, {\it Critical Mass: How One Things Leads to 
Anorther}, Arrow Books, 2006.

\bibitem{helbig}
D. Helbig, Rev. Mod. Phys. {\bf 73}, 1067-1141 (2001).

\bibitem{glomo}
X. Zeng, R. Bagrodia, M. Gerla, Proc. 12th Workshop on Parallel and 
Distributed Simulations (PADS 98), 154-161, 1998.

\bibitem{qualnet}
www.scalable-networks.com

\bibitem{vanet}
L. Bononi, M. Di Felice, G. D. Angelo, M. Bracuto, L. Donatiello, 
Computer Networks {\bf 52}, 155-179, 2008.

\bibitem{book-virus}
P. Szor, {\it The Art of Computer Virus Research and Defense}, Symantec Press, 2006. 

\bibitem{maziar-vanet}
M. Nekovee, Proc. IEEE Vehicular Technology Conferenc (VTC07-Spring),
2486-2490, 2007. 

\bibitem{allenandtildesley}
M. P. Allen and D.J. Tildesley, {\it Computer Simulations of Liquids}, 
Clarendon, Oxford, 1987.

\bibitem{fujimoto} 
R. Fujimoto, {\it Parallel and Distributed Simulation Systems
}, Wiley-Interscience,NY, 1999. 

\bibitem{conserve1}
Z. Ji, J. Zhou, M. Takai, J. Martin, R. Bogardia, Proc. 18th IEEE Workshop on 
Parallel and Distributed Simulations (PADS'04), 2004.

\bibitem{conserve2}
L. Bononi, M. Di Felice, M. Bertini, E. Croci, Proc. of MSWIM'06, 
Torremolinos, Spain, 28-35, 2004.

\bibitem{synch2} G. Korniss, M. A. Novotny, H. Guclu, Z. Toroczkai and P. A. 
Rikvold, Science {\bf 299}, 677 (2003).


\bibitem{sci-am}
M. Hypponen, Scientific American, 70--77, November 2006.

\bibitem{nature}
J. Kleinberg, Nature {\bf 449}, 287-288, 2007.

\bibitem{maziar-worms}
M. Nekovee, New J. Phys. {\bf 9}, 189, 2007. 
\bibitem{bio}
N. T. J. Bailey, {\it The Mathematical Theory of Epidemics},
Hafner Publishing, 1957.
\bibitem{wisec}
K. Hwang, Y-k Kwok, S. Song, M. C. Y. Chen, Y. Chen, R. Zhou and X. Lou,
GridSec: Trusted grid computing with security binding and self-defence against network worms and DDoS attacks, Proc. International Conference on Grid Computing Security and Resource  Management (GSRM'05), Atlanta, USA, may 2005.

\bibitem{mmp}
L. F. Perrone and D. M. Nicol, Proc. IEEE MASCOTS 2000.
\end{thebibliography}
\end{document}